\documentclass[superscriptaddress,pre,showpacs,twocolumn,aps]{revtex4-1}
\usepackage{psfrag}
\usepackage{amssymb,amsmath,amsthm,color}
\usepackage{graphicx}

\newcommand{\beq}{\begin{equation}}
\newcommand{\eeq}{\end{equation}}

\begin{document}
\title{Run-and-tumble in a crowded environment: persistent exclusion process for swimmers}

\author{Rodrigo Soto}
\affiliation{Departamento de F\'{\i}sica, Facultad de Ciencias F\'{\i}sicas y Matem\'aticas Universidad de Chile,
Av. Blanco Encalada 2008, Santiago, Chile}
\affiliation{Rudolf Peierls Centre for Theoretical Physics, University of Oxford, Oxford OX1 3NP, UK}

\author{Ramin Golestanian}
\affiliation{Rudolf Peierls Centre for Theoretical Physics, University of Oxford, Oxford OX1 3NP, UK}

\date{\today}

\begin{abstract}
The effect of crowding on the run-and-tumble dynamics of swimmers such as bacteria is studied using a discrete
lattice model of mutually excluding particles that move with constant velocity along a direction that is randomized
at a rate $\alpha$. In stationary state, the system is found to break into dense clusters in which particles
are trapped or stopped from moving. The characteristic size of these clusters predominantly scales as $\alpha^{-0.5}$
both in 1D and 2D. For a range of densities, due to cooperative effects, the stopping time scales as
${\cal T}_{1d}^{0.85}$ and as ${\cal T}_{2d}^{0.8}$, where ${\cal T}_d$ is the diffusive time associated
with the motion of cluster boundaries. Our findings might be helpful in understanding the early stages of biofilm formation.
\end{abstract}


\maketitle

\section{Introduction}

Bacterial biofilms are fascinating examples of biological development into multicellular communities through
aggregation \cite{biofilm-1}, and exhibit novel properties such as differentiation and delegation
of function \cite{biofilm-2}. A prominent feature of biofilms is that the bacteria inhabiting them are phenotypically different from their
free-swimming counterparts \cite{biofilm-3}. For example, when individual bacteria interact with surfaces,
they undergo transformations that help them adapt to the new conditions by adopting different modes
of motility \cite{Gerard}. A biofilm  could nucleate when a number of bacteria decide to settle down
near a surface and become completely localized. In addition to environmental cues, a key factor in triggering
such phenotypic transformations for individual bacteria is the time they spend in any given neighborhood
and under the influence of the corresponding local conditions; a factor that could be strongly influenced
by {\it crowding}. If they happen to be stuck in a place for a certain amount of time, they decide to adapt to
the new lifestyle by converting into the appropriate phenotype. Therefore, a natural question arises from
a physics viewpoint: what is the time motile bacteria could spend being held up in their own traffic jams
near a surface?

Self-propelled particles with excluded volume are known to undergo jamming transition at relatively high volume 
fractions \cite{fily,speck}. They  also develop jammed regions at low global concentrations
when they have a local alignment interaction \cite{Peruani06}. It has been argued that such density patterns could form
due to effective mobility reduction in the presence of nearby swimmers \cite{Kawasaki97,CatesSpinodal}. This reduction
has been experimentally characterized by measuring the self-diffusion process \cite{Kudrolli10}. Using a coarse-grained
description of the swimmer dynamics and considering a mean-field approximation for the mobility reduction, it has been
shown that the system phase separates into dense and dilute regions via a spinodal decomposition with long-time
coarsening \cite{CatesSpinodal}. If local ferromagnetic alignment is included, jammed clusters of macroscopic size
with internal grain boundaries are formed \cite{Peruani11}.

Here, we study a lattice model that mimics the bacterial run-and-tumble dynamics by allowing the particles to have
a persistent motion, and incorporates excluded volume by only allowing single occupancy for each site. 
The model was introduced  in Ref. \cite{Thompson}. In this paper, however, the hopping probability to occupied sites
is gradually decreased when the site density is increased by allowing a maximum occupancy of 100
particles in each site. In our work excluded volume is strictly enforced at each site and we show that this exclusion gives rise to new cooperative effects.
The resulting persistent exclusion process (PEP) is a lattice implementation of the continuous model presented in
Refs. \cite{Schnitzer93,CatesSpinodal,Rava}. 

For a wide range of particle concentrations and tumbling rates,
and both in 1D and 2D, the PEP generates clusters of jammed particles, separated by dilute regions with moving
particles. We probe the stationary distribution and the characteristic size of the clusters, the statistics of
the stopping time for individual particles, and the reduction in the average number of moving particles due
to jamming, as functions of the concentration and the tumbling rate. We find novel universal scaling behaviors,
and a counterintuitive effective superdiffusive process for the particles inside the jammed clusters
when the cluster boundary moves enough to reach them and open up the way for their release.

The plan of the paper is as follows. In Sec. \ref{sec.model} we describe the lattice model and present its main features. Simulation results for 1D and 2D are presented in Sec. \ref{sec.results}. Section \ref{clustersize1D} presents a derivation for the exponential distribution of cluster sizes in 1D and gives theoretical values for the characteristic  length. Finally, concluding remarks are given in Sec. \ref{sec.conclusions}

\section{Model} \label{sec.model}

We consider a lattice in one and two dimensions, consisting of $N$ sites, where each site can be occupied by at
most one particle. To model the persistent motion in the run phase, particles are characterized by a label that
indicates the direction they move: left or right in 1D and left, right, up, or down in 2D.  The dimensionless
particle concentration is $\phi$ and $M=\phi N$ is the total number of particles. At each time step $M$ particles
are chosen sequentially at random. For each particle, a new direction is chosen at random with probability $\alpha$
to model the tumble events; otherwise it preserves the previous direction. At this stage, if
the neighboring site as pointed by the director is empty, it will move to this new position. The evolution of
the model is asynchronous (sequential update) and stochastic. Our model does not include (ferromagnetic)
alignment interaction. Periodic boundary conditions are used. Units are chosen such that the lattice spacing
and the time step are fixed to unity. Note that at each step particles can be chosen more than once and on average $M$ particles are chosen. Therefore the particle motion is fluctuating with a unitary average speed. Finally, particles are initially placed at random mutually excluding positions and
each particle is assigned a random direction.

Our model is related to other lattice models used to describe nonequilibrium behavior, in certain limits.
When $\alpha=1$, the particles are reoriented at each time step and no persistent motion is obtained.
This corresponds to the equilibrium symmetric exclusion process (SEP) \cite{Huber1977,Richards1977}. The asymmetric exclusion process (ASEP) is a nonequilibrium process in
1D that breaks the detailed-balance by breaking the spatial symmetry, resulting in a directed flux \cite{ASEP,ASEPReview}.
This process and the totally asymmetric exclusion process (TASEP) show interesting nonequilibrium properties that
result from the particle interactions \cite{ASEP,ASEPReview,TASEP,Chou,Mallick,Parm}. In the PEP, the nonequilibrium
feature is introduced via the persistent motion that is controlled by $\alpha$; a property that is absent in the ASEP
and TASEP cases. If the limit of infinite dilution, each particle moves in straight-line run-segments for periods
of time that are geometrically distributed with the average of $\alpha^{-1}$, followed by a complete reorientation
(tumble). Consequently, at time scales longer than $\alpha^{-1}$, each particle performs a random walk with an
effective diffusion coefficient of $D_1=\alpha^{-1}$ \cite{berg1}. At finite concentrations, cooperative effects appear,
which are the main interest of this article.

\begin{figure}[t]
\includegraphics[width=.85\columnwidth]{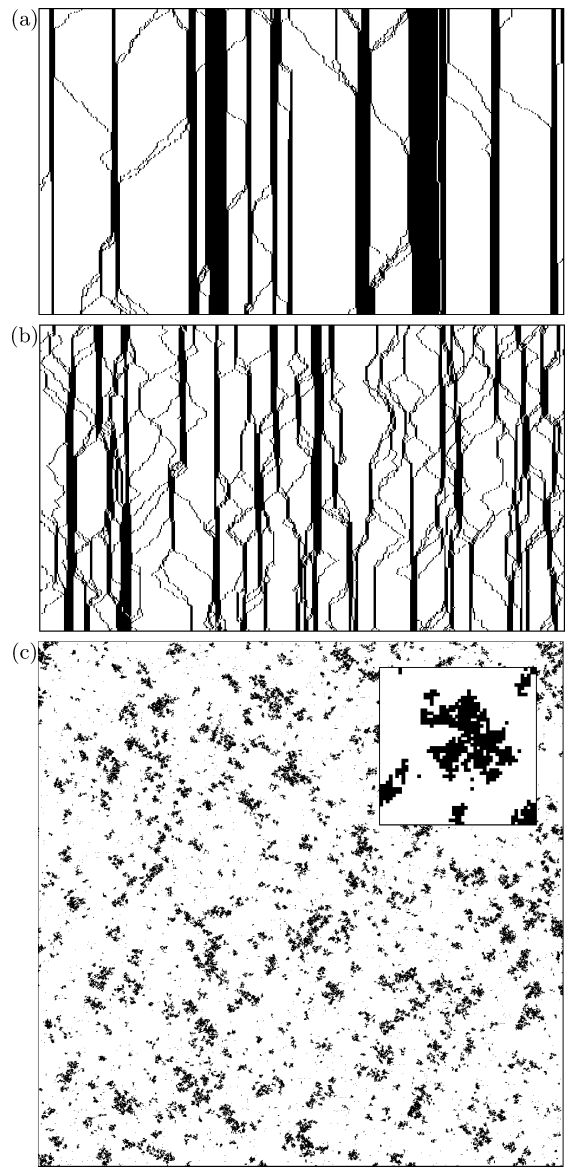}
\caption{(a) Spatiotemporal plot with $N=2000$, $\phi=0.2$, and $\alpha=0.01$ in 1D.
For clarity, only a fraction of the system $0\leq x\leq500$ is shown for a time
window of $T=300$ in the stationary regime, with the time increasing
upwards. The particles are shown as black dots. In the gas phase, they move with unit
average velocity but the stochastic dynamics produce fluctuating trajectories. (b) Same as in (a) with $\alpha=0.05$.
(c) Configuration of the stationary state in 2D with $N=1000\times1000$, $\phi=0.1$,
and $\alpha=0.001$. The particles are shown as black dots. Inset: Amplification
to show a single cluster.
}
\label{fig.snap}
\end{figure}

When $\alpha=0$ at any finite concentration, the system evolves to an absorbing state with all particles jammed.
Jamming is first produced by two particles that happen to move on the same track but in opposite directions, creating 
a cluster seed. Moving particles are absorbed into the existing clusters until no isolated
particles remain. At finite values of $\alpha$ the system forms clusters that coexist with dilute regions. This is a dynamic
state as the particles at the borders of the clusters can evaporate. The cluster dynamics involves three processes:
formation by the collision of two particles, absorption of moving particles at the boundaries, and evaporation
at the boundaries. No merging or splitting processes are possible as the clusters are immobile.

\section{Results} \label{sec.results}
\subsection{Stationary configurations}

We present results obtained in the stationary regime of simulations in 1D with $N=2000$ sites, spanning the parameter range
$0.01 \leq \phi \leq 0.9$ and $0.001 \leq \alpha \leq 1$, and the total number of time steps in the simulation is $T_{\rm max}=10^7$. In 2D the lattice
size is $N=1000 \times 1000$, with $0.01\leq \phi \leq 0.5$, $0.001 \leq \alpha \leq 1$, and $T_{\rm max}= 4 \times 10^6$.

Figures~\ref{fig.snap}a and \ref{fig.snap}b present spatiotemporal diagrams of the system in 1D. The system is clearly separated into dense clusters and dilute regions (denoted here as gas), which alternate.
Gas regions contain none or some isolated moving particles. Clusters, on the contrary, are close packed,
with the particle on the left (right) border pointing to the right (left). The interior particles have a distribution
of orientations as a result of successive reorientations they experience. This internal structure of the clusters
explains why we can observe rapid emission of two or more particles in the figure. When a border particle tumbles,
the neighboring particle could be pointing outward with probability $1/2$, and thus the two could be emitted
together in successive time steps. More particles could follow with decreasing probabilities. On average two
particles will be emitted in each evaporation process.

We can argue that the above picture holds when $\alpha\ll\phi$. If $\phi_g$ denotes the average particle
concentration in the gas phase, the cluster absorbs a gas particle at a rate $\phi_g/2$. 
Dynamic equilibrium
is reached when $\phi_g=\alpha$, since the evaporation rate is $\alpha/2$  as only half of the tumbling processes
are effective in pointing into the escape direction. Therefore, the cluster sizes are adjusted via the equilibration
with the gas phase. The described picture can only be valid if $\phi_g<\phi$, i.e. if $\alpha<\phi$.
Here, we  assumed that the gas phase is homogeneous, which is the case if the average size of the gas regions $l_g$ is such that the particle travel time is smaller than the time between particle emission at the borders (i.e. $l_g\ll\alpha^{-1}$). Using the expression for $l_g$ derived in Section \ref{clustersize1D} we obtain the condition $\alpha\ll\phi$. 
 The above simple
kinetic picture for the dynamic equilibrium suggests that the position of each cluster boundary when $\alpha\ll\phi$
undergoes a diffusive motion with diffusion coefficient $D_{\rm border}=\alpha$. This is to be contrasted with the single
particle diffusion coefficient $D_1=\alpha^{-1}$. In the opposite case of $\alpha > \phi$, the dynamics is no longer
only via the motion of the cluster boundaries, and we must consider the internal dynamics of the gas phase.

Figure \ref{fig.snap}c shows a snapshot of the configuration of the 2D system in stationary state, which shows
a similar pattern of clusters. As in the 1D case, we observe that even at low global area fractions, the system
reaches stationary states with an abundance of clusters and a few moving particles.

\subsection{Cluster sizes}

\begin{figure*}[htb]
\includegraphics[width=2\columnwidth]{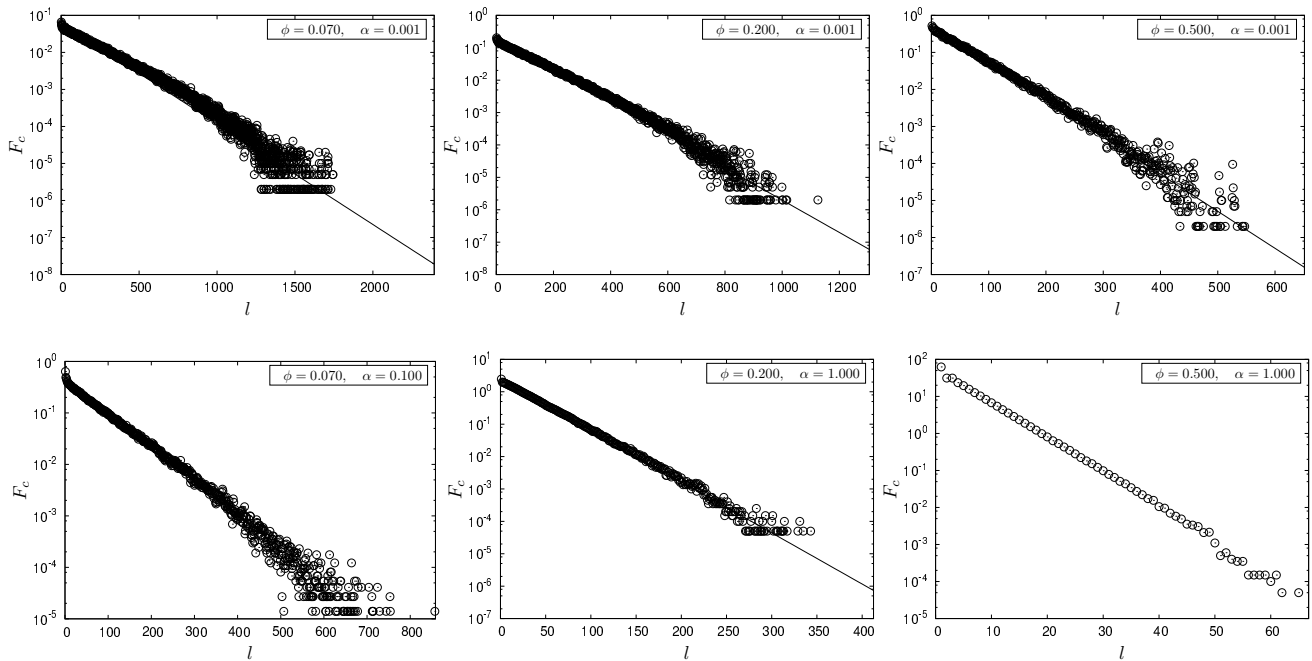}
\caption{Cluster size distribution in 1D for various concentrations $\phi$ and tumbling rates $\alpha$. The symbols are the results from the simulations and the solid line is an exponential fit.}
\label{fig.1D}
\end{figure*}

\begin{figure*}[t]
\includegraphics[width=2\columnwidth]{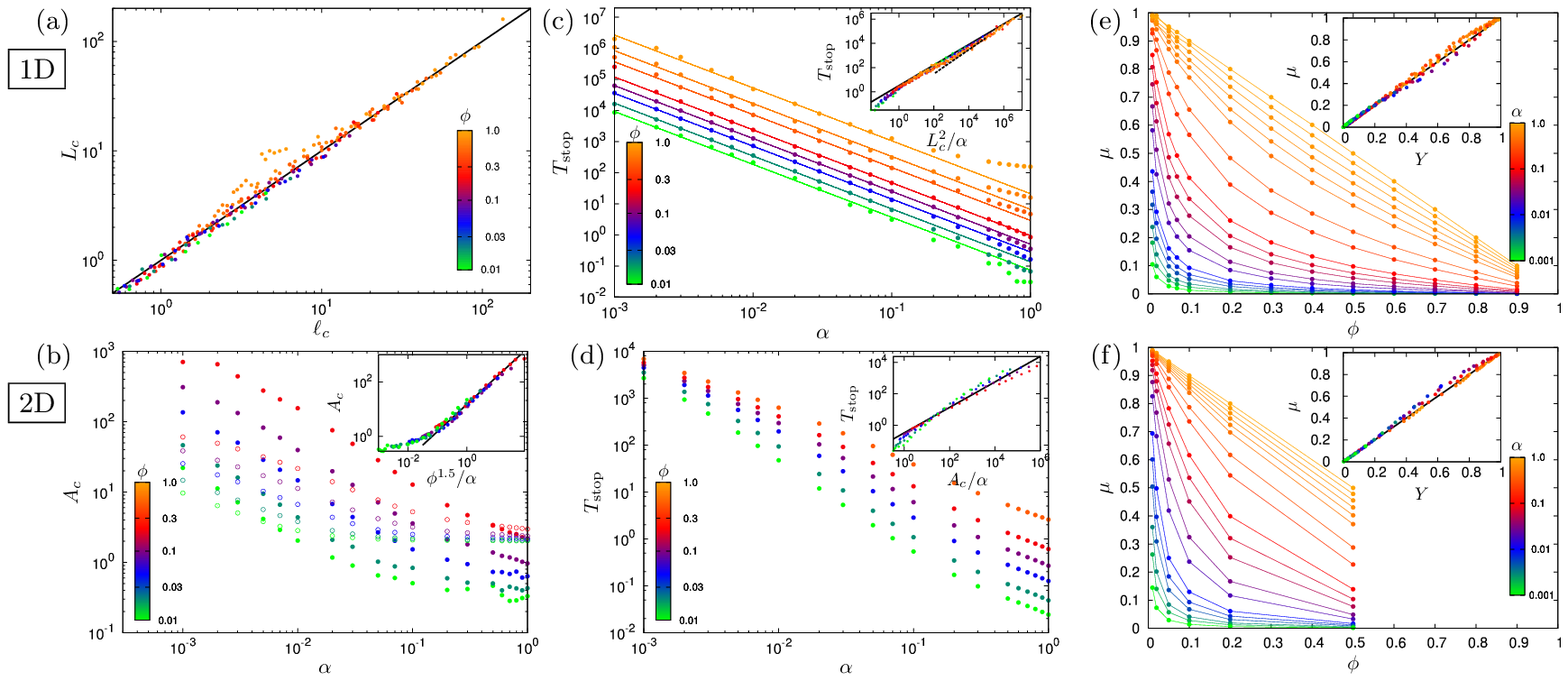}

\caption{(a) Data collapse plot for the average cluster size $L_c$ as a function of the length scale $\ell_c$ in 1D.
The points are obtained from simulations spanning the range $0.01\leq\phi\leq0.9$ and $0.001\leq\alpha\leq1$.
Concentrations are coded in color. The straight line corresponds to $L_c=\ell_c$.
(b) Characteristic cluster size (solid symbols) and average cluster size (open symbols) as a function of $\alpha$ in 2D.
Concentrations are coded in color. Inset: Data collapse of the characteristic cluster against $\phi^{1.5}/\alpha$.
(c) Average stopping time $T_{\rm stop}$ as a function of $\alpha$ in 1D (c) and 2D (d). Concentrations are coded in color. 
In (c), the straight lines are power law fits for each concentration with exponent $1.70 \pm 0.01$. Inset in (c): 
Data collapse plot against the time scale $L_c^2/\alpha$. The solid line is a power law fit with exponent 0.85 and 
the dashed line corresponds to exponent 1. Inset (d): Data collapse plot against the time scale $A_c/\alpha$. 
The solid line is a power law fit with exponent 0.8.
Fraction of moving particles $\mu$ as a function of $\phi$ in 1D (e) and 2D (f). Colors indicate the value of $\alpha$. 
Inset: Representative data collapse as a function of $Y_1=\alpha(1-\phi)(1+\phi+6\phi\alpha)/(\phi+\alpha+6\phi\alpha)$ 
in 1D (e) and $Y_2=\alpha(1-\phi)/[\alpha+0.6(1-\alpha)\phi]$ in 2D (f). The straight line corresponds to $\mu=Y$.
}
\label{fig.plot}
\end{figure*}

We find that in the stationary state the cluster sizes in 1D are exponentially distributed for a wide range of parameters, as shown in Fig. \ref{fig.1D}.
For small sizes the distributions differ from the exponential law as discussed in Sec. \ref{clustersize1D}.
The characteristic length scale $L_c$ depends on $\alpha$ and $\phi$. This dependence appears to be captured reasonably well
by the length scale $\ell_c=\sqrt{2\phi/[(1-\phi) \alpha]}$ that can is derived from a simple argument based
on maximizing configuration entropy in Sec. \ref{clustersize1D}, as the data collapse in
Fig.~\ref{fig.plot}a shows. 

\begin{figure*}[htb]
\includegraphics[width=2\columnwidth]{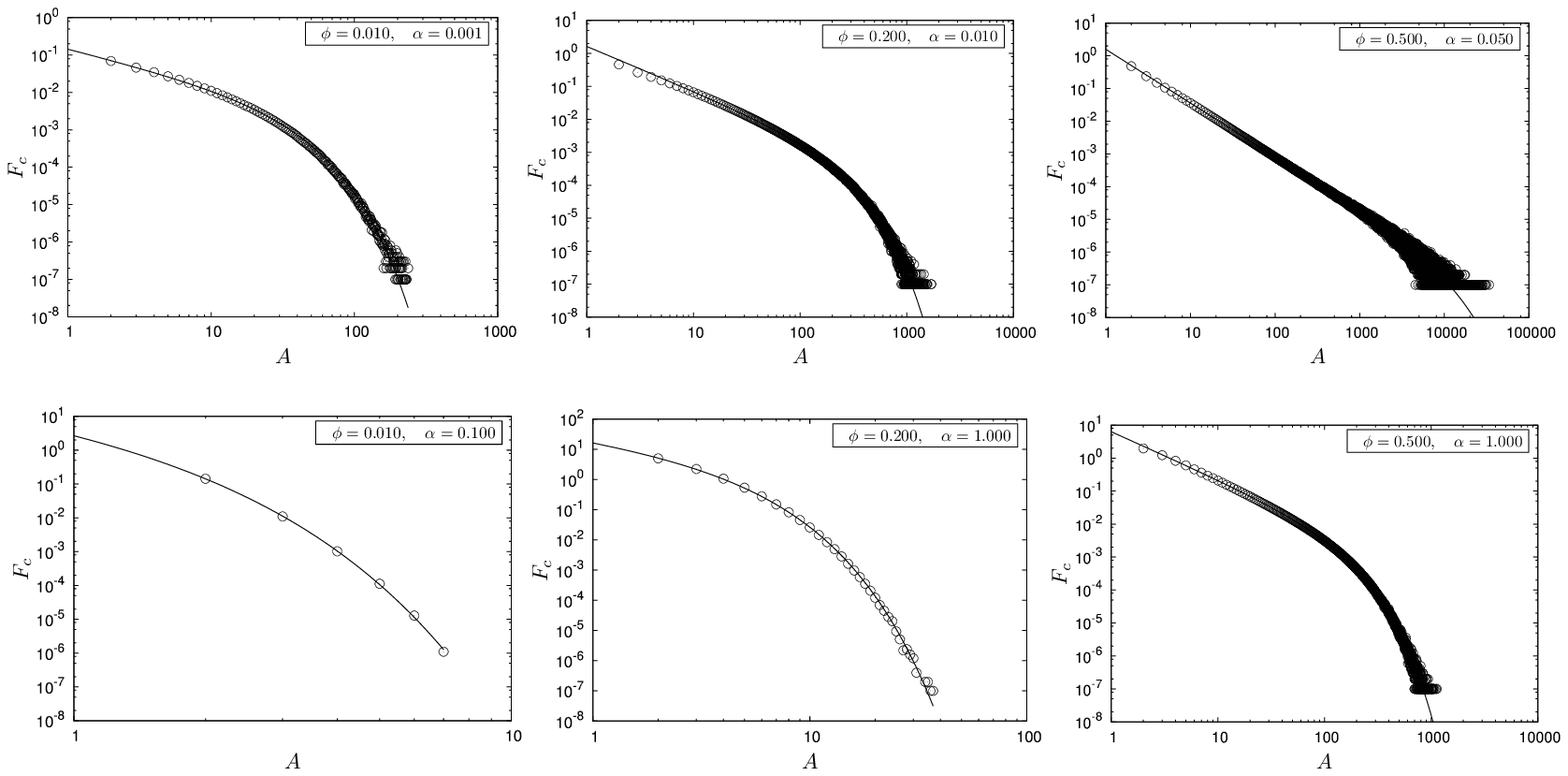}
\caption{Cluster size distribution in 2D for various concentrations $\phi$ and tumbling rates $\alpha$. The symbols are the results from the simulations and the solid line is  fit to the model $F_c = c A^{-\nu} e^{-A/A_c}$. The exponent $\nu$ lies in the range 1-1.5.}\label{fig.2D}
\end{figure*}

In 2D, the cluster areas $A$ are not distributed exponentially, but rather
as $A^{-\nu} e^{-A/A_c}$, where the exponent $\nu$ lies in the range $1$-$1.5$, as shown in Fig. \ref{fig.2D}. Figure \ref{fig.plot}b shows
the average $\langle A\rangle$ and the characteristic area $A_c$ as functions of $\alpha$ and $\phi$.
We find data collapse by plotting $A_c$ versus $\phi^{1.5}/\alpha$, with the asymptotic behavior of
$A_c\sim\phi^{1.5}/\alpha$ for large values. Note that in this limit $A_c$ has the same dependence
on $\alpha$ as $L_c^2$. On the contrary, the nonlinear dependence on $\phi$ indicates that cooperative
effects should be taken into account, and that the dynamics of the clusters is more complicated than
the 1D case with point boundaries. For example, we observe that clusters are not necessarily compact
in 2D; they might have holes as shown in the inset of Fig. \ref{fig.snap}c.

\subsection{Stopping time}

\begin{figure}[t]
\includegraphics[height=.6\columnwidth]{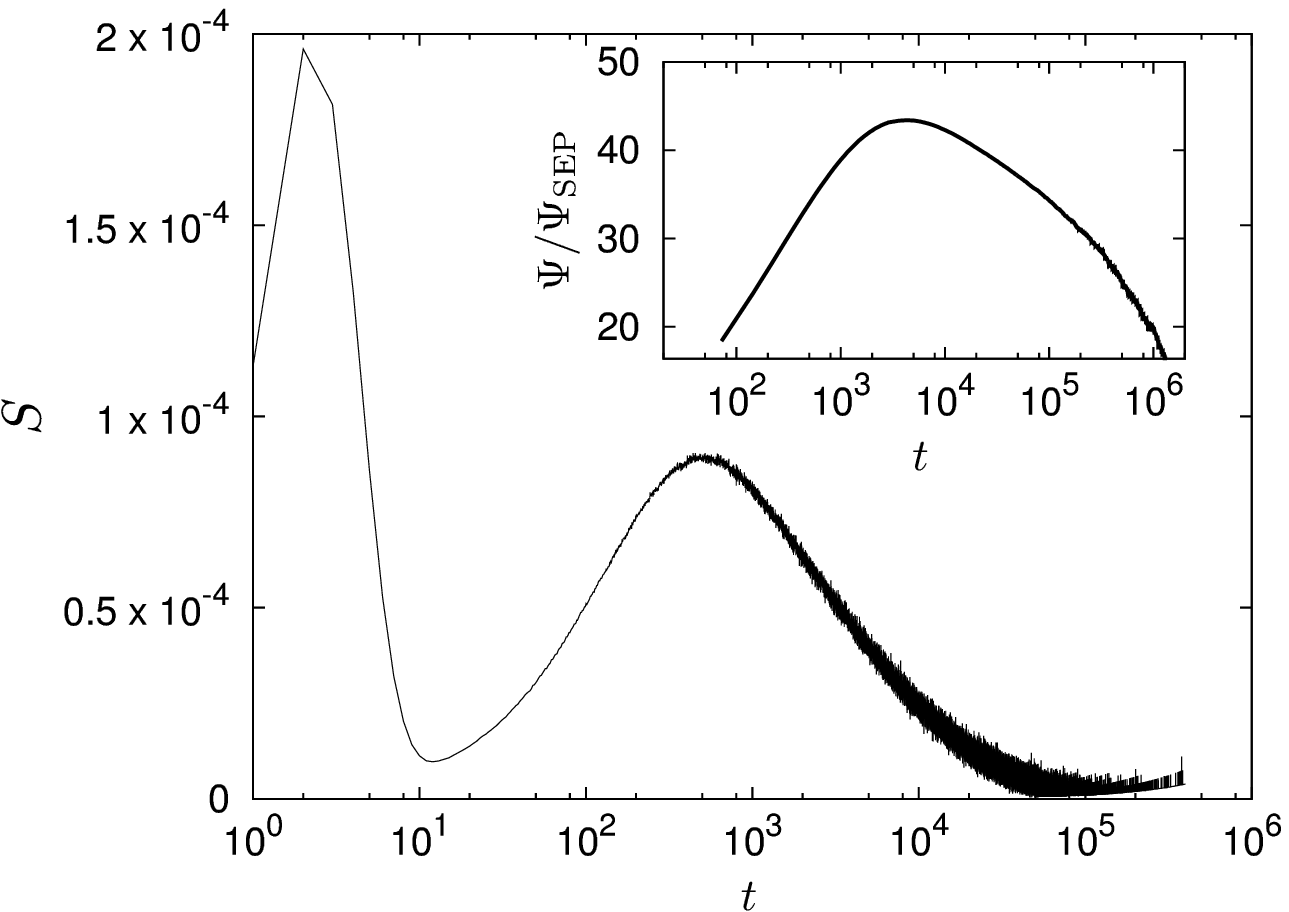}
\caption{Distribution of stopping times $S(t)$ (main figure) and overlap function divided by the SEP value (inset) for simulations in 1D with $\phi=0.5$ and $\alpha=0.005$.}
\label{fig.SPsi}
\end{figure}

Since the clusters are immobile on the whole, the particles in the clusters will be stopped until they
find a way to escape. When a moving particle joins a cluster it stops until it tumbles over a period
of $\alpha^{-1}$. If new particles arrive during this period, the particle will be moved to the interior
of the cluster and remain stopped for much longer times. We probe the statistics of the stopping times
and find that they are widely distributed with a prominent tail at long times. 
 The distribution of stopping times $S(t)$, where $S(t)$ is the average fraction of particles that have stopped
for $t$ time steps,  shows an interesting behavior. For large $\alpha$ it is dominated by particles with small stopping times, whereas for small $\alpha$ it is bimodal as shown in Fig.\ \ref{fig.SPsi} revealing two distinct populations with small and large stopping times. In the latter case, it is the population with large stopping times that dominate the dynamics.

The average stopping times, defined as $T_{\rm stop}=\sum_t t S(t)$  are presented in Fig.~\ref{fig.plot}c (1D) and Fig.~\ref{fig.plot}d (2D) as functions
of $\alpha$ for different values of $\phi$. In 1D, we observe a remarkable universal scaling behavior
$T_{\rm stop} \sim \alpha^{-1.7}$ over several orders of magnitude. One might naively expect
that the average stopping time is of the order of the time it takes for the cluster boundaries to diffuse
across the cluster size $L_c$, namely,
${\cal T}_{1d} \equiv L_c^2/D_{\rm border}\approx2\phi/[(1-\phi) \alpha^2]$. Using this combination in fact does lead
to data collapse for all relevant $\alpha$ and $\phi$ values [see Fig.~\ref{fig.plot}c inset].
However, the scaling plot has an effective exponent of 0.85, namely, $T_{\rm stop} \sim {\cal T}_{1d}^{0.85}$.
This means that the effective motion of the boundary is in fact superdiffusive, presumably due to cooperative
effects. In the 2D case, the stopping times presented in Fig. \ref{fig.plot}d do not show a uniform power law
dependence on $\alpha$ as in the 1D case. However, they do collapse to a good approximation when plotted against
the diffusive time scale ${\cal T}_{2d} \equiv A_c/\alpha$ [see Fig.~\ref{fig.plot}d inset]. As in the 1D case,
a superdiffusive behavior is found, which is consistent with an approximate power law behavior of
$T_{\rm stop} \sim {\cal T}_{2d}^{0.8}$ although the data collapse is less accurate than in the 1D case.

\subsection{Site-site temporal correlation}
To further characterize the dynamics we also study the site-site temporal correlation or overlap function in 1D. In 2D the computation of the overlap function is computationally expensive and we do not present results here.
At each site, we define $s_i$ to be zero if the site is empty and one otherwise, regardless of the director
of the particle. The overlap function is defined as
\beq
\Psi(t) = \frac{\langle s_i(t') s_i(t+t') \rangle - \phi^2 }{\phi-\phi^2},
\eeq
which has been adequately normalized to give $\Psi(0)=1$ and vanish in the absence of correlations.
The averaging is done over time $t'$ and site position.
The overlap function for the SEP in 1D has been
computed exactly resulting in $\Psi_{\rm SEP} = e^{-t} I_0(t)$, with the asymptotic large time
expression $\Psi_{\rm SEP}\sim 1/\sqrt{2\pi t}$, where $I_o$ is the modified Bessel function
of the first kind of zeroth order \cite{Huber1977,Richards1977}. Figure \ref{fig.SPsi} (inset) shows
the overlap function divided by $\Psi_{\rm SEP}$, such that the effect of persistence in
the dynamics can be probed. At intermediate times, we observe a large increase in the overlap
as compared to the SEP. This enhanced overlap, that increases for decreasing $\alpha$, is due
to the presence of jammed regions that trap particles for long times. At short times, the left
and right borders of a given cluster perform independent random walks, and the overlap function
exhibits an initial decay $\Psi=1-(1-\phi)\sqrt{2\pi t}$. At longer times, the clusters diffuse
and the particles can perform a diffusive exclusion process, recovering the decay $\Psi\sim1/\sqrt{t}$,
albeit with a larger amplitude than the SEP as an effect of the retention time in the jammed regions.
The crossover time between these two regimes, $T_c$, can be obtained numerically as the instant when
$\Psi/\Psi_{\rm SEP}$ is maximum. We find the relation $T_{\rm stop}=2 T_c$ between the two time scales
over the entire range of parameters studied, which shows that the two definitions are consistent
and probe the same phenomenon. That is, $T_{\rm stop}$ acts as a characteristic time scale for the dynamics of the 
system and controls the overlap correlations.

\subsection{Mobility}
At each time step, only a fraction $\mu$ of the particles can move and the others are jammed.
To understand the behavior of this quantity, which we call mobility, as a function of $\phi$
and $\alpha$, it is helpful to examine its limiting forms. 

When $\alpha\approx1$, we expect
to obtain the SEP mobility that equals the fraction of empty sites, i.e. $\mu_{\rm SEP}=1-\phi$  \cite{Huber1977,Richards1977}. 
It should decrease as $\alpha$ is decreased, and vanish when $\alpha=0$. When $\phi$ is increased beyond $\alpha$, we expect the relative
mobility $\widehat{\mu}=\mu/(1-\phi)$ to decrease, due to the formation of the jammed regions.
Besides the limit $\lim_{\alpha\to1} \widehat\mu=1$, other known limits are 
$\lim_{\alpha \to 0}\widehat\mu = 0$, and $\lim_{\phi \to 0}\widehat\mu =1$. Note that the $\phi=0$
and $\alpha=0$ limits do not conmute. The simulations show that $\lim_{\phi \to 1}\widehat\mu$ is a smooth function
of $\alpha$ that interpolates between 0 and 1.

The measured values of mobility are presented in Fig.~\ref{fig.plot}e (1D) and Fig.~\ref{fig.plot}f (2D)
as functions of $\phi$ for different values of $\alpha$.We observe that the mobility is reduced more
strongly than linearly with respect to $\phi$ as the tumbling rate is decreased. We have used cutoff
densities of 0.9 in 1D and 0.5 in 2D, where glassy behavior is not present. We find that it is possible to collapse the results
for $\mu$ versus $\phi$ and $\alpha$, by using functions that interpolate smoothly between the 
known asymptotic limits, chosen from the class of low-rank rational functions. 
Figure \ref{fig.plot}e (inset) shows one such collapse plot using a representative
interpolation function  $Y_1=\alpha(1-\phi)(1+\phi+6\phi\alpha)/[\phi+\alpha+6\phi\alpha]$ for the 1D
case. For the 2D case, Fig. \ref{fig.plot}f (inset) shows one such collapse plot using
a representative interpolation function $Y_2=\alpha(1-\phi)/[\alpha+0.6(1-\alpha)\phi]$.
We note that this mobility is the average value over the system where the clusters and the gas
regions coexist, and it should not be confused with a mobility that depends on
the local concentration \cite{CatesSpinodal,Kudrolli10}.

\section{Derivation of the cluster size distribution in 1D} \label{clustersize1D}
When $\alpha\ll\phi$, the cluster borders in 1D move by evaporation and absorption processes. Assuming that clusters do not interact, the probability distribution of the right border position $R$ is described by a discrete-time master equation with transition rates $W_{R\to R+1}=\phi_g/2$ and $W_{R\to R-1}=\alpha/2$.
The stationary state is reached when the particle density is such that the transition rates balance, leading to $\phi_g=\alpha$. 
At the stationary state the drift terms vanish, resulting in the master equation for the symmetric random walk
\begin{eqnarray}
P(R,t+1) &=& \left[1-\alpha\right] P(R,t) \nonumber\\
&&+ \frac{\alpha}{2} \left[P(R+1,t) + P(R-1,t)\right] 
\end{eqnarray}
Therefore, each border performs a diffusive motion with diffusion coefficient $D_{\rm border}=\alpha$.
Note that, as a result of the particle interactions, this is is contrast with the single particle diffusion coefficient $D_1=\alpha^{-1}$.

To obtain the cluster size distribution in 1D and the crossover time scale we note that as the left and right borders of a cluster perform independent random walks, the cluster size also performs  a random walk. If we assume that the cluster interaction is weak and that the relevant interaction is through the uncorrelated emission and absorption of gas particles, then each cluster evolves independently of others, except for the particle conservation constraints. This can be mapped to an equilibrium process for the sizes of independent pseudo-particles and, hence, the size distribution can be obtained by the maximization of a relevant configurational entropy. 
The particle concentration in the gas phase is $\phi_g$ fixes the total number of particles in the cluster phase, $N_{c}$.
This restriction is imposed by a Lagrange multiplier $\lambda$ and a second multiplier $\gamma$ is also included to fix the total number of clusters $C$. Then, the number of clusters of size $l$, $F_c(l)$, is obtained from the maximization of the entropy $S = \log[C!/\prod_l F_c(l)!] -\lambda[N_{c} - \sum_l l F_c(l)] - \gamma[C-\sum_l F_c(l)]$,
which yields an exponential distribution 
\beq
F_c(l) = A_c e^{-l/\ell_c}.
\eeq
A similar reasoning allows us
to derive the distribution of sizes in the gas regions as $F_g(l) = A_g e^{-l/\ell_g}$.

The constants $A_c$, $A_g$, $\ell_c$, and $\ell_g$ can be fixed with the following prescriptions. (i)  The average sizes are such that the total concentration is recovered:
$\langle l_c\rangle\phi_c + \langle l_g\rangle \phi_g=  (\langle l_c\rangle + \langle l_g\rangle )\phi$.
(ii) The gas and cluster regions cover the entire system: $\sum_l l F_c(l) + \sum_l l F_g(l) = N$.
(iii) There is equal number of gas and cluster regions: $\sum_l F_g(l) = \sum_l F_c(l)$.
(iv) The dynamical balance in the number of dimers: Normally, the number of gas particles
in each gas region is smaller than one, and the production of dimers is due to
the emission in a time interval smaller than $\ell_g$ of gas particles on the two sides
of a gas region. Their production rate is then $W_2^+\approx \alpha(1-e^{-\alpha \ell_g/2})\sum_l F_g(l)$.
Their decay rate is simply due to evaporation with a rate $W_2^-\approx \alpha F_c(2)$. The final condition
is then $W_2^+=W_2^-$. 

Assuming that the cluster length scale $\ell_c$ is large and approximating
the sums by integrals, the transcendental equations can be solved to give 
\begin{eqnarray}
\ell_c&\approx& \sqrt{2 \phi/[(1-\phi) \alpha]},\\
\ell_g&\approx&\sqrt{2(1-\phi)/(\phi \alpha)}.
\end{eqnarray} 
Both predicted length scales diverge when $\alpha\to 0$,
and their behavior with the volume fraction is opposite, with $\ell_c$ diverging at close packing.

As shown in Fig. \ref{fig.1D} the distributions follow exponential laws except for small clusters because in these cases the dynamics is fast and we cannot consider anymore that their border obey equilibrium random walk processes.

The derivation made for the exponential distribution in 1D is not valid in 2D. As it is shown in Fig.\ 1c, clusters in 2D are not compact objects (they have voids and they are not circular in shape) and, therefore, they are not simply caracterized by their area $A$ (number of particles belonging to the cluster). More parameters are necessary to fully describe them.
Also, in 2D it is not possible to make the mapping to an Brownian process as it was done in 1D when $\alpha\ll\phi$.

\section{Concluding remarks} \label{sec.conclusions}

We have examined the combined effect of crowding and run-and-tumble dynamics in an idealized model of bacterial 
dynamics and identified novel power law behaviors. Further work is needed to incorporate other aspects of bacterial 
interactions with solid surfaces and among themselves---including hydrodynamic effects, chemical signaling, 
and short-range interactions---to develop a better physical understanding of the early stages of biofilm formation.

The stopping time is a local probe for single bacteria to make decisions about phenotypic adaption to their 
local environment. The stopping time can be represented, for sufficiently long times, 
via a scaling form of $T_{\rm stop} \sim G(\phi) \alpha^{-b}$, with $G_1(\phi)=[\phi/(1-\phi)]^{0.85}$ 
and $b_1=1.7$ for the 1D case, and $G_2(\phi)=\phi^{1.2}$ and $b_2=1.6$ in 2D.
The simple form can be used as a rule-of-thumb expression for simple estimates and generic observations. 
First, it suggests that the clusters that could serve as the seeds needed 
for the early stages of biofilm formation could occur at relatively small densities without the need of long-ranged communication or coordination between bacteria, provided the tumbling rate is sufficiently 
small. 
Moreover, the stopping time is significantly more sensitive to the tumbling rate than it is 
to the density of the bacteria that form the crowd, despite the fact that exclusion due to crowding is 
the cause of stopping. Interestingly, the chemotaxis circuitry provides a sensitive control over the tumbling 
rate for bacteria, which means that modulation of the tumbling rate could provide a feedback mechanism 
related to the formation of immobile clusters. Eventual strategies to either reduce or to increase the tumbling rate once in a crowded environment should be possible to measure experimentally. 

\acknowledgments

This research is supported by Fondecyt Grant No. 1100100 and Anillo grant ACT 127 (R.S.),
and Human Frontier Science Program (HFSP) grant RGP0061/2013 (R.G.).

\end{document}